\documentclass[useAMS,usenatbib]{mn2e}

\usepackage{caption}
\usepackage{epsfig}
\usepackage{graphicx}
\usepackage{rotating}
\usepackage{textcomp}
\usepackage{latexsym}
\usepackage{varioref}
\usepackage{xspace}
\usepackage{makeidx}
\usepackage{verbatim}
\usepackage{tabularx}

\title[Hard X-ray mCVs]
{Hard X-ray properties of magnetic cataclysmic variables\thanks{Mainly based on observations with \textit{INTEGRAL}, an ESA project with instruments and science data centre funded by ESA member states (especially the PI countries: Denmark, France, Germany, Italy, Spain), Czech Republic, and Poland, and the participation of Russia and the USA.}}
\author[S. Scaringi \textit{et al.}]
{S. Scaringi$^{1}$\thanks{E-mail: simo@astro.soton.ac.uk}, A.J. Bird$^{1}$, A.J. Norton$^{2}$, C. Knigge$^{1}$, A.B. Hill$^{1}$\thanks{New address: Laboratoire d'Astrophysique de Grenoble, UMR 5571, Universit\'{e} Joseph Fourier, BP 53, 38041 Grenoble, France}, D.J. Clark$^{1}$,\newauthor A.J. Dean$^{1}$, V.A. McBride$^{1}$, E.J. Barlow$^{2}$, L. Bassani$^{3}$, A. Bazzano$^{4}$, M. Fiocchi$^{4}$ \newauthor and R. Landi$^{3}$\\ $^{1}$Department of Physics and Astronomy, University of Southampton, Highfield, Southampton, SO17 1BJ, UK \\ $^{2}$Department of Physics and Astronomy, The Open University, Walton Hall, Milton Keynes, MK7 6AA, UK\\ $^{3}$INAF$-$IASF Bologna, Via P. Gobetti 101, I$-$40129 Bologna, Italy\\ $^{4}$INAF$-$IASF Rome, Via Fosso del Cavaliere 100, I$-$00133 Roma, Italy\\}

\begin{document} 

\date{}

\pagerange{\pageref{firstpage}--\pageref{lastpage}} \pubyear{2009}

\maketitle

\label{firstpage}

\begin{abstract}
Hard X-ray surveys have  proven remarkably efficient in detecting intermediate polars and asynchronous polars, two of the rarest type of cataclysmic variable (CV). Here we present a global study of hard X-ray selected intermediate polars and asynchronous polars, focusing particularly on the  link between hard X-ray properties and spin/orbital periods. To this end, we first construct a new sample of these objects by cross-correlating candidate sources detected in \textit{INTEGRAL}/IBIS observations against catalogues of known CVs. We find 23 cataclysmic variable matches, and also present an additional 9 (of which 3 are definite) likely magnetic cataclysmic variables (mCVs) identified by others through optical follow-ups of IBIS detections. We also include in our analysis hard X-ray observations from \textit{Swift}/BAT and \textit{SUZAKU}/HXD in order to make our study more complete. We find that most hard X-ray detected mCVs have $P_{spin}/P_{orb}<0.1$ above the period gap. In this respect we also point out the very low number of detected systems in any band between $P_{spin}/P_{orb}=0.3$ and $P_{spin}/P_{orb}=1$ and the apparent peak of the $P_{spin}/P_{orb}$ distribution at about $0.1$. The observational features of the $P_{spin} - P_{orb}$ plane are discussed in the context of mCV evolution scenarios. We also present for the first time evidence for correlations between hard X-ray spectral hardness and $P_{spin}$, $P_{orb}$ and $P_{spin}/P_{orb}$. An attempt to explain the observed correlations is made in the context of mCV evolution and accretion footprint geometries on the white dwarf surface.
\end{abstract}

\begin{keywords}
gamma-rays: observations, surveys, X-ray binaries, magnetic cataclysmic variables
\end{keywords}

\section{Introduction}
Cataclysmic variables (CVs) are close binary systems consisting of a late-type star transferring material onto a white dwarf (WD) via Roche-lobe overflow. Magnetic CVs (mCVs) are a small subset of the catalogued CVs ($\approx 10\% - 20\%$, \citealt{downes}; \citealt{RKcat}), and fall into two categories: polars (or AM Her types after the prototype system) and intermediate polars (IPs or DQ Her types). The WDs in polars possess such strong magnetic fields that they can synchronise the whole system, yielding $P_{orb}=P_{spin}$. The strong magnetic field in these systems is confirmed by strong optical polarisation and measurments of cyclotron humps (\citealt{warner}). Accretion in polars is thought to follow the magnetic field lines of the WD straight from the L1 point onto the WD magnetic poles, and no accretion disk is expected (for a review of polars, see \citealt{cropper}). Another possible class of systems called asynchronous polars (APs) might exist, where the spin and orbital periods are out of synchronisation by only a few percent. It is not known exactly why this is but one suggestion is that these systems are polars which have had a recent nova event, kicking them slightly out of synchronisation (\citealt{warner}). For IPs, the lack of strong optical polarisation implies a weaker magnetic field, not powerful enough to synchronise the secondary (for a review of IPs, see \citealt{patterson}). In these systems, material leaving the L1 point usually forms an accretion disc up to the point where the magnetic pressure exceeds the ram pressure of the accreting gas. From this point onwards the accretion dynamics are governed by the magnetic field lines, which channel the material onto the WD magnetic poles. The nature of these systems is confirmed by the detection of coherent X-ray modulations associated with the spin period of the WD.

In the simplest scenario for X-ray production in mCVs, the magnetically channeled accretion column impacts the WD poles producing hard X-rays from thermal bremsstrahlung cooling by free electrons with kT of the order of 10s of keV  (\citealt{warner, cropper}). The emission is thought to originate in the post-shock region, a region below the shock front created from the impacting accretion column. Softer X-rays are also produced from the absorption and reprocessing of these higher energy photons in the WD photosphere. As a result, both polars and IPs are expected to emit high energy photons, but discrepancies exist between the observed ratio of soft-to-hard X-rays between polars and IPs, with polars showing an excess of observed soft X-rays (\citealt{lamb}).  \cite{chanmugan} and others have reported that the total X-ray luminosities of IPs are greater than those of polars by a factor of $\approx 10$, attributed mainly to the higher accretion rates. Moreover it has been proposed that strong magnetic fields in polars produce a more ``blobby'' flow than in IPs (\citealt{warner}). These high density ``blobs'' are then able to penetrate within the post-shock region, emitting fewer bremsstrahlung photons and contributing more to the observed X-ray blackbody spectral component, thought to be produced at the base of the post-shock region.

In recent years an increasing number of mCVs have been detected and discovered by hard X-ray telescopes such as {\it INTEGRAL}/IBIS (\citealt{barlow,landi}), {\it Swift}/BAT (\citealt{brunsch}) and {\it SUZAKU}/HXD (\citealt{terada}). Here, we present a newly compiled list of IBIS mCVs, forming part of a larger compilation of hard X-ray selected IPs from all three telescopes mentioned above, allowing us to study the general properties of this hard X-ray selected sample.

This selected sample of CVs will allow us to investigate the observational properties of the $P_{orb}$--$P_{spin}$ plane, and compare them to mCV evolution models. Moreover, we will investigate the spectral properties of our sample, and present for the first time observed spectral hardness correlations observed within IPs as a function of their orbital and spin parameters.

\section{Recent Hard X-ray observations}

The \textit{INTEGRAL} satellite, launched in October 2002, has now carried out more than 6 years of observations in the energy range 5 keV to 10 MeV. In particular, the \textit{INTEGRAL}/IBIS survey is one of the main mission objectives. The IBIS detector (\citealt{ubertini03,lebrun03}) has been optimised for survey work, with a large field of view ($30^o$) and with unprecedented sensitivity in the soft-gamma ray regime, yielding excellent imaging capabilities. It is worth emphasising that the IBIS survey has been optimised to detect faint persistent sources, which mCVs are. The aim of the survey is to expand the current knowledge of the 20-100 keV sky by cataloguing high-energy sources and examining their properties, both individually and globally. The IBIS survey dataset consists of dedicated observations along the Galactic plane and around the Galactic centre. Additionally, a combination of pointed and deep exposure observations are added to the dataset once they become public. As a result, the latest release of the \textit{INTEGRAL}/IBIS survey provides all-sky coverage, albeit with spatially variable sensitivity. The depth of the IBIS survey has increased significantly with each release (\citealt{bird1st,bird2nd,bird07}) and has now reached a peak sensitivity corresponding to a flux limit below 1 mCrab in the 20-100 keV range. The latest study on \textit{INTEGRAL} CVs by \cite{landi} contained 22 sources found within the IBIS/ISGRI survey or identified as CVs later on through optical follow-up.  

This work complements that of \cite{landi} and \cite{barlow}, in particular including a larger dataset yielding deeper observations. The IBIS dataset used consists of over 36,000 individual Science Window (ScW) pointings, covering 6 years of observations (revs $\approx$ 46-660). These have been processed with the latest pipeline (\textit{OSA 7.0}, \citealt{goldwurm03}), and mosaics were created from the deconvolved images in 5 energy ranges\footnote{20-40keV, 30-60keV, 20-100keV, 17-30keV and 18-60keV}. Staring and performance verification (PV) observations have not been used for the mosaic creation, and noisy ScWs have been excluded based on the RMS of the individual images. These new survey maps are a considerable improvement on any previous ones, due to the new pipeline software, together with the significantly increased exposure times.

\textit{Swift}, has been optimised to locate gamma-ray bursts, and as a consequence the main hard X-ray instrument, the Burst Alert Monitor (BAT), has similar capabilities to IBIS, possessing a large field of view and operating in essentially the same energy range. BAT has also been used for survey work (\citealt{tueller}), and in particular has also detected a high number of IPs (\citealt{brunsch}). In order to make our study as complete as possible, we have decided to include the IPs from \cite{brunsch} observed with the {\it Swift}/BAT detector. Moreover, one extra IP has been added, AE Aqr, as observed by \textit{SUZAKU}/HXD (\citealt{terada}), sampling again a very similar energy range to IBIS. The similarity of energy range allows us to construct a hard X-ray selected sample with minimal biases.

Prior to this work, hard X-ray detections have been reported for $\approx 30$ mCVs above 10 keV. More than $90\%$ of these are IPs, and there are also two rare asynchronous polars. When compared to the older soft X-ray selected samples of mCVs the picture is somewhat different. First of all, as one would expect, soft X-ray detectors are more sensitive to mCVs and consequently will produce larger samples; including equal amounts of polars and IPs. 

\section{The Hard X-ray CV population}

\subsection{\textit{INTEGRAL}/IBIS CVs}
Here we describe the catalogue matching procedure adopted in order to identify known CVs in our IBIS dataset. To do this we require the most complete and up-to-date catalogue of such objects, thus we merge the two most complete CV catalogues in the literature: The Catalogue and Atlas of Cataclysmic Variables (\citealt{downes}, hereafter DWScat) and the Catalogue of Cataclysmic Binaries (\citealt{RKcat}, hereafter RKcat). DWScat contains 1830 CVs whilst RKcat contains 731, and we note that 656 CVs are common to both catalogues. The principal reason for RKcat having fewer objects is that only CVs with known orbital periods are included in the sample; however, RKcat also includes a few CVs that DWScat does not report. Our final known CV set therefore contains 1905 CVs (hereafter DRKcat). Fewer than $\approx 9\%$ of the total number of CVs within DRKcat are known to be magnetic in nature (included in the catalogue as either DQ Her, AM Her or IP), and only approximately $3\%$ (56 sources) are IPs.

Catalogue matching has been performed between the total CV set produced, which contains 1905 CVs and a preliminary IBIS candidate excess list containing real sources constructed in the same way as for the ISINA algorithm (\citealt{scaringi}, Section 3), but now on a larger dataset, containing over 9000 excesses constructed from the public data available for revolutions 46 -660. This includes excesses detected in the final mosaics, revolution mosaics and revolution sequence mosaics in any of the 5 main energy bands in order to locate variable candidates too. 

We have carried out the same correlation exercise as in \cite{barlow}, using a similar method to \cite{stephen}. Similarly to \cite{barlow} we employ a search radius of $4^\prime$, which also corresponds well with the expected error on faint IBIS detections (\citealt{gros}). For a search radius of $4^\prime$ we obtain 56 sources as confirmed or candidate CVs, of which 23 are expected to be false coincidences. We have visually inspected all of the correlated sources and found 33 matches coincide with mainly non-CV globular cluster sources and previously identified X-ray objects; however some are image artefacts related to the Galactic centre region. It is important to note that with a $\approx 2^\prime$ typical source location accuracy it is very hard to associate a detection with an optical counterpart alone. We have performed the same exercise by increasing the search radius to $5^\prime$ which increases the sample to inspect to 76 candidates. We find that all are false matches with the possible exception of the Dwarf Nova DN V1830 Sgr located $4.8^\prime$ away from the IBIS detection detected in revolution 106 (MJD 53128.8 - 53131.7). This is slightly outside the $90\%$ error radius for a $\approx 6 \sigma$ detection and we cannot definitely associate the two at the moment.

Table \ref{drk-ibis} shows the main characteristics of the 23 objects identified from our correlation analysis. We have estimated distances for this sample using the method described by \cite{knigge}, based on the evolution, of the donor star where 2MASS K-band magnitudes were available (\citealt{2MASS}) and $P_{orb} < 6.2$ hours\footnote{The limit for which the method is applicable}. In addition we show in Table \ref{extra-cvs} the 9 other IBIS-detected mCVs used in this work. These were not part of the correlation analysis, because they were not present in DRKcat, but have been identified through optical spectroscopy following the IBIS discovery. Of the 23 objects considered to be real matches from our analysis, 17 are previously known \textit{INTEGRAL} detected CVs, whilst 6 sources are new detections. Most of the new objects are of the intermediate polar subclass with the possible exception of TW Pic, which is considered by some as a possible VY Scl star (\citealt{norton_b}), and AX J1832.3$-$0840, which is not identified in full at the moment.

In total this paper presents 32 IBIS-detected CVs, increasing the number from the latest previous study of \cite{landi} by 10 systems. 

\begin{sidewaystable*}\small
\caption{Results of the IBIS $-$ DRK catalogue matching with $4^\prime$ search.}
\label{drk-ibis}
\begin{center}

\begin{tabular}{ c c c c c c c c c c c c c}
\hline
Name\footnote{$\star$ indicates new hard X-ray detections} & $\alpha,\delta$\footnote{Right ascension and declination in degrees, J2000} & type\footnote{IP=intermediate polar, P=polar, AP=asynchronous polar, N=nova, DN=dwarf nova. All are confirmed except for $\bullet$:probable, $\bullet\bullet$: possible} & offset\footnote{Angular distance between the DRKcat catalogue positions and the IBIS coordinate} & Map code\footnote{IBIS detection only. Map with maximum significance: (B1) 20-40 keV; (B2) 30-60 keV; (B3) 20-100 keV; (B4) 17-30 keV; (B5) 18-60 keV; in brackets appears the significance value.} & Count rate\footnote{Determined between 20-100 keV} & Exposure & Flux\footnote{The flux is calculated assuming a power law spectra with index of 2.9, the average index for IPs (\citealt{barlow})} & $P_{orb}$ & $P_{spin}$ & Distance\footnote{The distances have been computed with 2MASS K band magnitudes (\citealt{2MASS}) using the method presented by \cite{knigge} based on the evolution of the secondary.} & Refs \\[0.5ex]
  & (IBIS position) &  & ($^\prime$) &  & ($ct~s^{-1}$) & (ks) & 20-100 keV & (min) & (s) & (pc) & \\[0.5ex]
\hline
1RXS J002258.3$+$614111 & 5.739,61.714 & IP & 1.7 & B4(8.7) & $0.15\pm0.01$ & 3767 & 0.81 & 241.98 & 563.53 & 510 & [1,2,3,4]\\
V709 Cas & 7.207,59.303 & IP & 0.8 &  B5(54.3) & $1.03\pm0.01$ & 3562 & 5.53 & 320.4 & 312.77 & 300 & [1,2,5]\\
XY Ari$\star$ & 44.047,19.457 & IP & 1.1 & B5(5.5) & $0.53\pm0.12$ & 119 & 2.85 & 363.884 & 206.298 & 610 & [1,2,6] \\
GK Per & 52.777,43.928 & IP/DN & 1.7 & B5(4.7) & $0.26\pm0.07$ & 277 & 1.4 & 2875.4 & 351.34 & - & [1,2] \\
TV Col$\star$ & 82.357,-32.819 & IP & 0.1 & B4(11.6) & $0.68\pm0.08$ & 248 & 0.37 & 329.181 & 1911 & 330 & [1,2,7,8]\\
TW Pic$\star$ & 83.766,-57.998 & IP?/VY Scl?$\bullet\bullet$ & 2.5 & B1(5.8) & $0.3\pm0.07$ & 363 & 1.61 & - & - & - & [1,2,9,10,11]\\
BY Cam & 85.728,60.842 & AP & 1.2 & B5(5.1) & $0.69\pm0.11$ & 162 & 3.7 & 201.298 & 11846.4 & 140 & [1,2]\\
MU Cam & 96.316,73.567 & IP & 0.6 & B4(5.4) & $0.24\pm0.06$ & 548 & 1.29 & 283.104 & 1187.24 & 440 & [1,2,12]\\
SWIFT J0732.5$-$1331$\star$ & 113.13,-13.513 & IP & 1.6 & B3(6.1) & $0.39\pm0.06$ & 409 & 2.09 & 336.24 & 512.42 & - & [1,2,13]\\
V834 Cen & 212.260,-45.290 & P & 0.9 & B1(5.4) & $0.16\pm0.03$ & 1675 & 0.86 & 101.51712 & 6091.0272 & 70 & [1,2]\\
IGR J14536$-$5522 & 223.421,-55.394 & P & 2.0 & B4 (11.9) & $0.27\pm0.03$ & 2658 & 1.45 & 189.36 & 11361.6 & 140 & [1,2,14]\\
NY Lup & 237.052,-45.481 & IP & 0.5 & B5(49.1) & $1.17\pm0.03$ & 3141 & 6.28 & 591.84 & 693.01 & - & [1,2,15]\\
V2400 Oph & 258.173,-24.279 & IP & 2.2 & B1(33.4) & $0.68\pm0.02$ & 4453 & 3.65 & 204.48 & 927.6 & 180 & [1,2,16]\\
1H 1726$-$058 & 262.606,-5.984 & IP & 0.7 &  B5(22.8) & $0.85\pm0.04$ & 1449 & 4.56 & 925.27 & 128 & - & [1,2,17]\\
V2487 Oph & 262.960,-19.244 & IP/N$\bullet\bullet$ & 2.3 & B3(9.1) & $0.18\pm0.02$ & 4562 & 0.97 & - & - & - & [1,2,18]\\
AX J1832.3$-$0840$\star$ & 278.083,-8.721 & ? & 3.1 & B4(5.5) & $0.07\pm0.03$ & 3090 & 0.38 & - & - & - & [1,2,19,20]\\
V1223 Sgr & 283.753,-31.153 & IP & 0.8 & B5(52.2) & $1.45\pm0.03$ & 2358 & 7.79 & 201.951 & 746 & 150 & [1,2]\\
V1432 Aql & 295.052,-10.421 & AP & 0.2 & B5(10.8) & $0.69\pm0.07$ & 429 & 3.7 & 201.938 & 12150.4 & 240 & [1,2,21,22]\\
V2069 Cyg & 320.906,42.279 & IP$\bullet$ & 1.8 & B5(6.2) & $0.21\pm0.03$ & 1648 & 1.13 & 448.824 & 743.2 & - & [1,2,23,24]\\
1RXS J213344.1$+$510725 & 323.446,51.122 & IP & 0.3 & B5(25.8) & $0.65\pm0.03$ & 2207 & 3.49 & 431.568 & 570.82 & - & [1,2,25]\\
SS Cyg & 325.698,43.582 & DN & 0.6 & B5(23.0) & $0.7\pm0.03$ & 1674 & 3.76 & 396.1872 & - & - & [1,2,26]\\
FO Aqr & 334.514,-8.354 & IP & 1.7 & B4(6.1) & $0.65\pm0.2$ & 54 & 3.49 & 290.966 & 1254.45 & 250 & [1,2,27]\\
AO Psc$\star$ & 343.815,-3.194 & IP & 1.3 & B4(4.8) & $0.43\pm0.11$ & 108 & 2.31 & 215.461 & 805.2 & 200 & [1,2,28,29]\\
\hline
\end{tabular}
\\

References: [1]\cite{RKcat}; [2]\cite{downes}; [3]\cite{bonnet_a}; [4]\cite{masetti_c}; [5]\cite{bonnet_b}; [6]\cite{norton_a}; [7]\cite{hellier_a}; [8]\cite{august}; [9]\cite{buckley_a}; [10]\cite{chen}; [11]\cite{norton_b}; [12]\cite{araujo}; [13]\cite{butters_b}; [14]\cite{masetti_a}; [15]\cite{demartino_a}; [16]\cite{hellier_b}; [17]\cite{gansicke}; [18]\cite{hernanz}; [19]\cite{muno}; [20]\cite{sugi}; [21]\cite{watson}; [22]\cite{geck}; [23]\cite{thor}; [24]\cite{demartinoatel}; [25]\cite{bonnet_c}; [26]\cite{friend}; [27]\cite{marsh}; [28]\cite{kalu}; [29]\cite{vaname}; 
\end{center}
\end{sidewaystable*}

\begin{sidewaystable*}\small	
\begin{centering}
\caption{Additional mCVs detected by IBIS not included in DRKcat.}

\begin{tabular}{c c c c c c c c c c c}
\hline
 Name & $\alpha,\delta$\footnote{Right ascension and declination in degrees, J2000} & type\footnote{IP=intermediate polar, P=polar, AP=asynchronous polar, N=nova, DN=dwarf nova. All are confirmed except for $\bullet$:probable, $\bullet\bullet$: possible} & Map code\footnote{IBIS detection only. Map with maximum significance: (B1) 20-40 keV; (B2) 30-60 keV; (B3) 20-100 keV; (B4) 17-30 keV; (B5) 18-60 keV; in brackets appears the significance value.} & Count rate\footnote{Determined between 20-100 keV} & Exposure & Flux\footnote{The flux is calculated assuming a power law spectra with index of 2.9, the average index for IPs (\citealt{barlow})} & $P_{orb}$&$P_{spin}$ & Distance\footnote{The distances have been computed with 2MASS K band magnitudes (\citealt{2MASS}) using the method presented by \cite{knigge} based on the evolution of the secondary.} & Refs \\[0.5ex]
  &  &  &  & ($ct~s^{-1}$) & (ks) & 20-100 keV & (min) & (k) & (pc) &\\[0.5ex]
\hline
IGR J08390$-$4833 & 129.705,-48.524 & IP$\bullet\bullet$  & B5(6.3) & $0.08\pm0.02$ & 3072 & 0.43 & -- & -- & -- & [1]\\
XSS J12270$-$4859\footnote{Note that there are conflicting results regarding the nature of this object. RXTE data suggests an IP nature, whilst \textit{SUZAKU} data suggests an LMXB nature. See references.} & 187.004,-48.906 & IP$\bullet\bullet$ & B5(9.9) & $0.42\pm0.04$ & 955 & 2.26 & -- & -- & -- & [2,3,4]\\
IGR J15094$-$6649 & 227.406,-66.823 & IP & B5(11.0) & $0.18\pm0.03$ & 3265 & 0.97 & 353.40 & 809.42 & 600 & [3,5]\\
IGR J16167$-$4957 & 244.140,-49.974 & IP$\bullet\bullet$ & B4(20.7) & $0.4\pm0.02$ & 3466 & 2.15 & -- & -- & -- &[3]\\
IGR J16500$-$3307 & 252.491,-33.114 & IP/DN & B1(13.3) & $0.33\pm0.03$ & 3353 & 1.77 & 217.02 & 597.92 & 270 & [5,6]\\
IGR J17195$-$4100 & 259.906,-40.997 & IP & B5(23.3) & $0.54\pm0.02$ & 4279 & 2.9 & 240.30 & 1139.55 & 120 & [2,3,5]\\
IGR J18173$-$2509 & 274.353,-25.158 & IP$\bullet\bullet$  & B1(14.9) & $0.27\pm0.01$ & 5744 & 1.45 & -- & -- & -- & [7]\\
IGR J18308$-$1232 & 277.696,-12.532 & IP$\bullet\bullet$  & B5(7.1) & $0.18\pm0.03$ & 3265 & 0.97 & -- & -- & -- & [8]\\
IGR J19267$+$1325 & 291.670,13.425 & IP$\bullet$  & B5(7.5) & $0.15\pm0.03$ & 2963 & 0.81 & -- & -- & -- &[9,10] \\
\hline 
\end{tabular}
\\ 
All classifications are correct except for, $\bullet$: probable classification $\bullet\bullet$: possible classification.
References:[1]\cite{knia}; [2]\cite{butters_a}; [3]\cite{masetti_a}; [4]\cite{saitou}; [5]\cite{retha09}; [6]\cite{masetti_b}; [7]\cite{masetti_VII}; [8]\cite{parisi}; [9]\cite{steeghs}; [10]\cite{evans}.
\label{extra-cvs}
\end{centering}
\end{sidewaystable*}

\subsection{\textit{Swift}/BAT and \textit{SUZAKU}/HXD CVs}

\textit{Swift}/BAT has also observed a large number of mCVs, and it is worth including these in our study for completeness. We decided to include all the 22 BAT detected IPs (\citealt{brunsch}), where 14 have been observed by IBIS as well. Similarly to the IBIS-only IPs, the BAT-only IPs are all placed above the period gap with the exception of EX Hya.

\textit{SUZAKU}/HXD on the other hand has a different observing strategy composed of small field of view pointings. This does not allow the telescope to produce survey data like IBIS or BAT, however mCVs have been detected and observed with HXD. In particular HXD had observed the IP AE Aqr (\citealt{terada}), which has not been observed with either IBIS or BAT, and therefore is included in our study as well. We caution however that the possible origin of the hard X-rays in AE Aqr could be non-thermal (\citealt{terada}). This however has not been shown in full, and given the power-law model fit to the 3-25 keV spectra of AE Aqr yielding an index of 2.10 (\citealt{terada}) not far from the indeces found in fast spinning mCVs (\citealt{landi}) we believe this object should still be included in our analysis.  

All the IPs observed with BAT and HXD used in this study are presented in Table \ref{table:BATHXD}, together with their orbital and spin periods. 

\begin{table*}
\begin{centering}
\caption{Additional IPs used in this work detected with \textit{Swift}/BAT and \textit{SUZAKU}/HXD with known spin and orbital periods.}
\begin{tabular}{c c c c c c c}
\hline
 Name & $\alpha,\delta$ & type & Detection & $P_{orb}$&$P_{spin}$  & Refs \\[0.5ex]
  &  &  &  & (min)& (s)  &\\[0.5ex]
\hline
V1062 Tau & 75.615,24.756 & IP & BAT & 597.60 & 3726 & [1]\\
TX Col & 85.834,-41.032 & IP & BAT & 343.15 & 1909.7 & [1]\\
V405 Aur & 89.897,53.896 & IP & BAT & 249.12 & 545.455 & [1]\\
BG CMi & 112.871,9.940 & IP & BAT & 194.04 & 847.03 & [1]\\
PQ Gem & 117.822,14.740 & IP & BAT & 310.80 & 833.4 & [1]\\
DO Dra & 175.910,71.689 & IP & BAT & 238.139 & 529.31 & [1]\\
EX Hya & 193.102,-29.249 & IP & BAT & 98.257 & 4021.62 & [1]\\
AE Aqr & 310.038,-0.871 & IP & HXD & 592.785 & 33.0767 & [2]\\
\hline 
\end{tabular}
\\ 
References:[1]\cite{brunsch}; [2]\cite{terada}.
\label{table:BATHXD}
\end{centering}
\end{table*}

\section{Hard X-ray selected magnetic CV properties}

As revealed in \cite{barlow} and \cite{landi}, the vast majority of IBIS detected CVs are magnetic (the only exception being SS Cyg). The total number of objects with known spin and orbital periods in our sample is 30 systems, of which 22 are IBIS detections (18 IPs). It is interesting to note the relative incidence of polar systems and intermediate polars. Optically selected samples favour the former, with the number of known polars being twice that of IPs. However, in our hard X-ray selected sample, the picture is very different, with only 4 polars being included in the sample (2 of which are APs). This result is expected, as IPs are known to produce $\approx 10$ times more hard X-rays than polars due to their higher mass transfer and  intrinsically harder spectrum (\citealt{warner}). 

Only two synchronous polars have been detected in our sample, both at relatively close distance and in regions of the sky with high exposure times. This leads us to conclude that the flux in the hard X-ray range for these systems is much lower when compared to IPs or APs. We note from Table \ref{drk-ibis} that the two detected polars are among the closest objects in our sample, and hence it is probably for this reason that IBIS (and BAT, \citealt{tueller}) can see them. We have also checked this by inspecting where other polars sit in the \textit{INTEGRAL} exposure map and conclude that more deep observations are required before more of these systems are detected in the hard X-ray range.

Two of the four confirmed asynchronous polars (APs) are also included in our sample. At first one might think that these systems should have properties resembling the polar class. However, as shown by \cite{schwarz}, their accretion rates are $\approx 10-20$ times greater than that of polars. Moreover both \textit{INTEGRAL} detected APs are 2-3 magnitudes brighter than the two non-detected APs in the K-band. This, together with our distance estimates suggests that our efficiency at detecting APs in the hard X-ray range is $\approx50\%$, similar to the IPs.

\subsection{The $P_{orb} - P_{spin}$ plane}

Theoretical simulations on the evolution of mCVs have been performed by \cite{norton_theo1,norton_theo2}. One interesting prediction of these models is that of different accretion flows for IPs depending on where they are in their evolutionary stage. In particular these models also predict the existence of spin equilibria among IPs, one at about $P_{spin}/P_{orb} \approx 0.1$ and a second one at $P_{spin}/P_{orb} \approx 0.6$ (depending on mass ratio).

\begin{figure*}
\centering
\includegraphics[width=\textwidth]{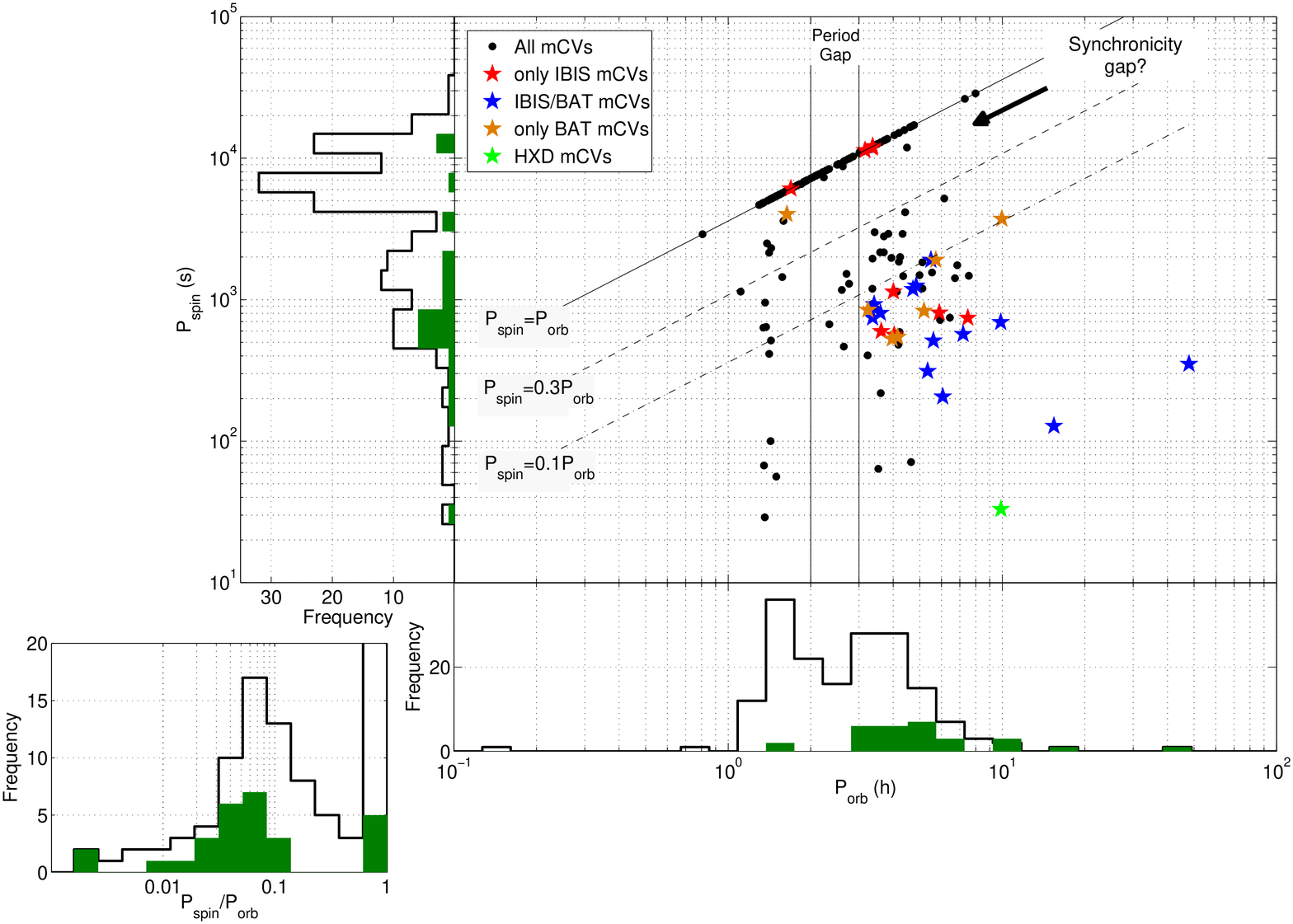}	
\caption{$P_{orb}$ vs. $P_{spin}$ for mCVs taken from RKcat. Stars indicate mCVs detected at hard X-ray energies. Also plotted is the period gap and ``synchronicity'' lines. For reference we also display the orbital, spin and synchronicity distributions. The shaded green areas represent hard X-ray selected mCVs.}
\label{fig:Po_Ps}
\end{figure*}

Figure \ref{fig:Po_Ps} shows the $P_{orb} - P_{spin}$ plane for mCVs together with their orbital, spin and synchronicity ($P_{spin}/P_{orb}$) distributions. The hard X-ray detected systems used in this work are shown with stars. Also plotted for reference is the period gap and three ``synchronicity'' lines showing $P_{spin}=P_{orb}$ (polars), $P_{spin}=0.1 P_{orb}$ and $P_{spin}=0.3 P_{orb}$. It is clear that most of the hard X-ray detected systems are above the period gap and have $P_{spin}\leq0.1P_{orb}$. The only IP system outside this range is EX Hya which is closer than 100 parsec (\citealt{brunsch}). Similarly to the two detected polars, it is of no real surprise that hard X-ray detectors can see this close object.

The fact that no IP has yet been observed with hard X-ray telescopes above the period gap with $P_{spin}/P_{orb}>0.1$ suggests that these IPs have different accretion flows compared to IPs with $P_{spin}/P_{orb}<0.1$. This idea is supported by the fact that the distribution of all known mCVs does indeed seem to peak at about $P_{spin}/P_{orb}\approx0.1$ in the bottom left panel in Fig.\ref{fig:Po_Ps}, where a spin equilibrium has been predicted. As IPs evolve from low synchronicity ($P_{spin}/P_{orb}<<0.1$) towards their $P_{spin}/P_{orb} \approx 0.1$ equilibrium, their accretion flows resemble those of propeller systems, where a lot of the material incoming from the secondary is actually propelled away and does not reach the pole of the WD. So in essence most IPs with $P_{spin}/P_{orb} << 0.1$ have yet to reach their equilibrium (regardless of field strength), and their accretion flows are different from any other IP elsewhere in the $P_{spin} - P_{orb}$ plane (\citealt{norton_theo1, norton_theo2}). As a consequence one would not necessarily expect the hard X-ray emission mechanisms to be the same for all IPs. 

Another interesting observational feature of this plane is that only one unconfirmed IP system, V697 Sco, lies within what we call the synchronicity gap: a region in the $P_{orb} - P_{spin}$ plane within $P_{spin}>0.3P_{orb}$ and total synchronicity above the period gap. The low number of IP systems with high synchronicity above the period gap is also predicted by the models. As explained by \cite{norton_theo1,norton_theo2}, as mCVs evolve, both their mass ratios and orbital period decrease. These trends individually cause opposite shifts in the spin-to-orbital ratio at which the spin-equilibrium occurs for a given magnetic field. As a result, typical IPs with magnetic field strength of a few MG will evolve from being disc-like accretors at long orbital period (where  $P_{spin}/P_{orb} \approx 0.1$), to ring-like accretors at short orbital period (where $P_{spin}/P_{orb} \approx 0.6$), providing that they do not synchronise along the way and become polars. If the magnetic field of the systems is of the order of a few hundred MG, on the other hand, then they will be prone to synchronise and become polars, crossing the synchronicity gap fairly quickly, helping explain the low number of systems in this region. This is the likely history of EX Hya and systems neighboring in the $P_{orb} - P_{spin}$ plane.

We therefore predict that, given the already long exposure times accumulated with IBIS and BAT, the IPs within $P_{spin}=0.1P_{orb}$ and $P_{spin}=0.3P_{orb}$ region will not be detected in significant numbers. Conversely we expect most of the IPs with $P_{spin}<0.1P_{orb}$ to be observable with longer exposure times and better sensitivity. We also do not expect many systems to exist within the synchronicity gap.

\subsection{Are hard X-ray detected mCVs different?}
It should be clear by now that hard X-ray telescopes are more sensitive at detecting the IP population rather than the polar one. However it is not yet clear whether hard X-ray telescopes are producing populations which are consistent with being drawn from the general mCV population. In order to assess this we have performed a Kolmogorov-Smirnov test (KS test) on all the $P_{spin}$, $P_{orb}$ and $P_{spin}/P_{orb}$ distributions of hard X-ray selected samples versus the known mCV population taken from RKcat. In all cases the test rejects the null hypothesis that the distributions are drawn from the same parent with $>99.99\%$ confidence.

As mentioned before, the difference in distributions within $P_{orb}$ (bottom panel in Fig. \ref{fig:Po_Ps}) between these sets could be caused by the fact that all mCVs below the period gap are intrinsically less luminous given their lower accretion rates. However it does not exclude the possibility that the X-ray emission mechanisms for mCVs below the period gap is substantially different than the mCVs above the gap. Both of these effects can be regarded as systematic, however the possibility of hard X-ray missions detecting homogeneous IP samples has already been suggested by \cite{gansicke}, and here we bring forward more observational evidence for this. It is clear however that if mCVs below the period gap emit hard X-rays, then current telescopes are not sensitive enough to detect them. 

Moreover, from the $P_{spin}/P_{orb}$ KS test (bottom left panel in Fig. \ref{fig:Po_Ps}) result we can also comment on the fact that hard X-ray telescopes are not sensitive to high synchronicity systems. These do not necessarily have to live below the period gap, and in fact about half of the polar population lives above the gap. This suggests that hard X-ray surveys are very insensitive to the polar population as well as mCVs below the period gap. We can also look at this result and suggest that, on the other hand, hard X-ray surveys are extremely sensitive to IPs with long $P_{orb}$ (i.e. above the period gap) and to systems with low $P_{spin}/P_{orb}$.

It is hard to say whether mCVs below the period gap and polars in general have different emission mechanisms with a KS test alone. However we believe that the evidence arising from the orbital and synchronicity distributions suggest this very strongly.

\section{Hardness plane correlations}

The production of hard X-ray photons from mCVs is thought to originate in the post-shock region of the WD by bremsstrahlung cooling of free electrons. This is somewhat different to softer X-rays ($<2$ keV) seen from mCVs, which can originate from a blackbody component close to the WD surface. In fact in recent years many medium resolution X-ray spectra have been obtained for different kinds of mCVs (\citealt{masetti_b,anzolin,schwarz,butters_a,evans07,done,landi}) and have been fitted with a soft blackbody component ($kT \approx 80 eV$) plus a hard component characterised by the stratified accretion column of \cite{cropper99}. For those mCVs that are detected in the hard X-ray range the detection only samples the bremsstrahlung component. In particular the hard X-ray  energy range ($>17$ keV) is telling us about the temperature distribution of components within the multi-temperature bremsstrahlung emission, not the ratio of hard-to-soft X-ray components.

\begin{figure}
\centering
\includegraphics[width=0.5\textwidth, height=7cm]{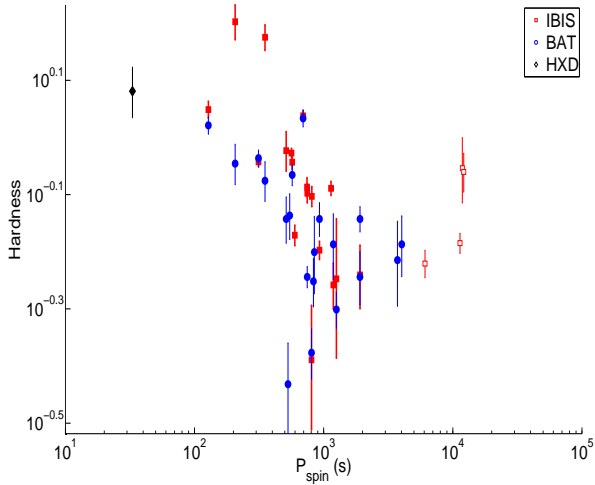}	
\caption{30-60/17-30 keV hardness versus spin period for the hard X-ray selected mCVs used in this work. Polars and APs are shown in empty squares.}
\label{fig:Pspin}
\end{figure}

\begin{figure}
\centering
\includegraphics[width=0.5\textwidth, height=7cm]{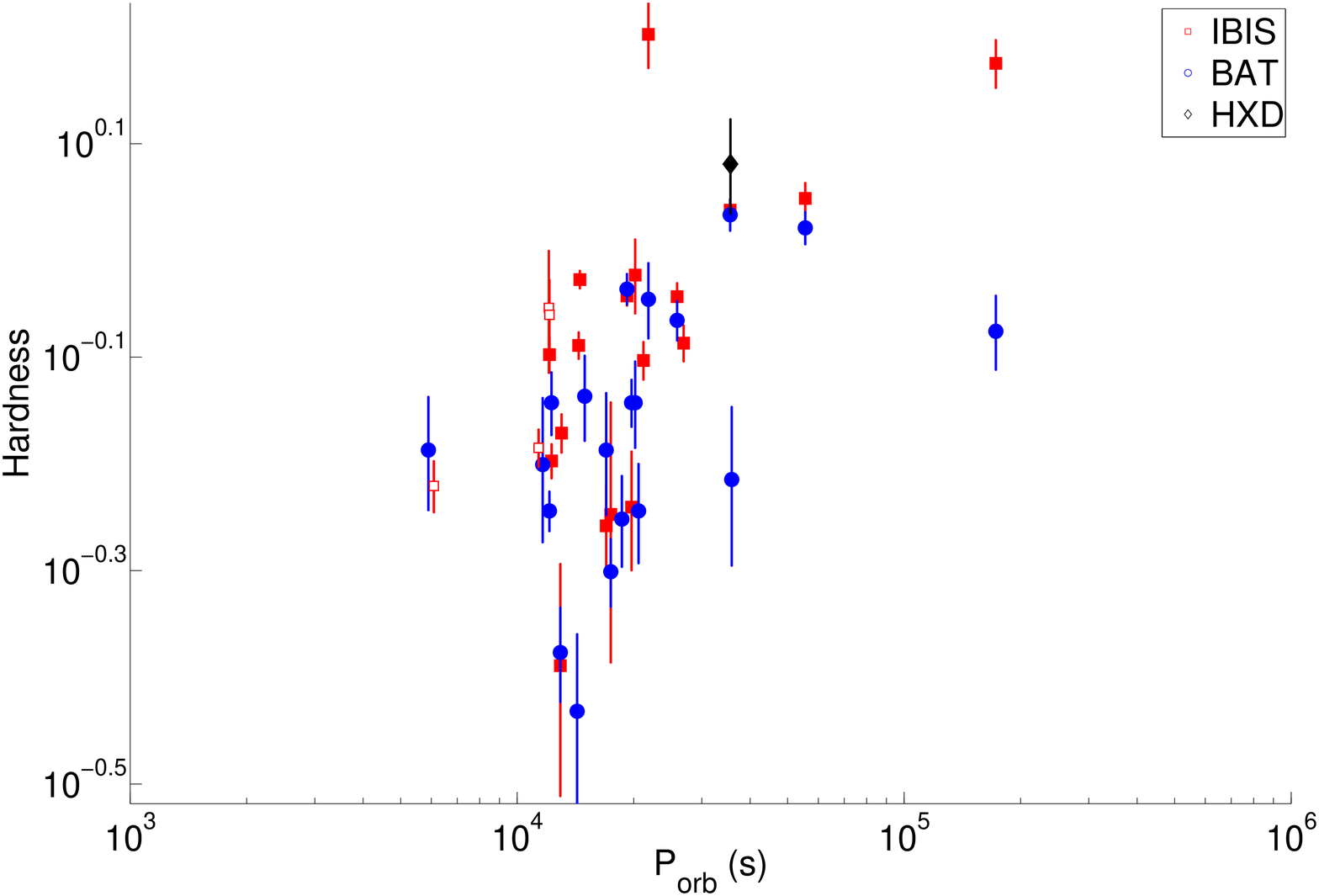}	
\caption{30-60/17-30 keV hardness versus orbital period for the hard X-ray selected mCVs used in this work. Polars and APs are shown in empty squares.}
\label{fig:Porb}
\end{figure}

\begin{figure}
\centering
\includegraphics[width=0.5\textwidth, height=7cm]{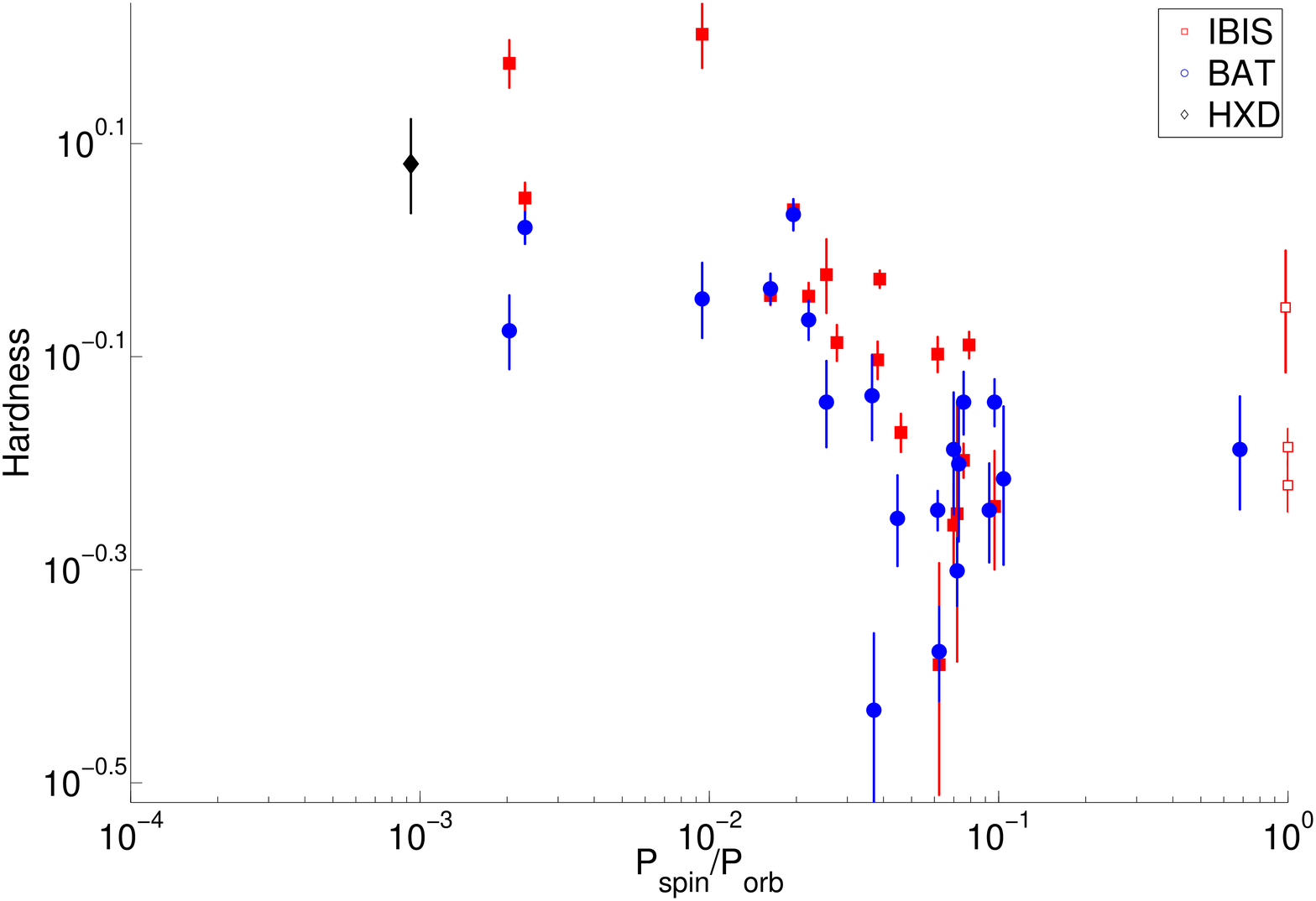}	
\caption{30-60/17-30 keV hardness versus $P_{spin}/P_{orb}$ for the hard X-ray selected mCVs used in this work. Polars and APs are shown in empty squares. We note the evident correlation for IPs with $P_{spin}/P_{orb} \leq 0.1$ }
\label{fig:Psync}
\end{figure}

\begin{figure}
\centering
\includegraphics[width=0.5\textwidth]{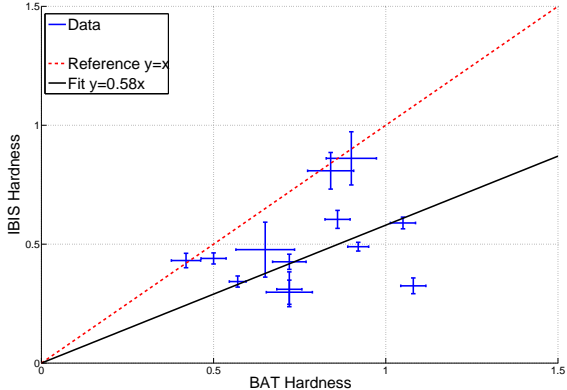}	
\caption{BAT extrapolated flux hardness vs. IBIS measured hardness. In blue is the data. In red we show a one to one line for reference. The cross-calibration fit is displayed in black. We note that both BAT and IBIS errors were taken into account when fitting.}
\label{fig:crosscalib}
\end{figure}

Keeping this in mind, Fig. \ref{fig:Pspin}, \ref{fig:Porb} and \ref{fig:Psync} shows the scatter plots for hardness, defined as the count ratio in the 30-60 keV and 17-30 keV bands, versus $P_{spin}$, $P_{orb}$ and $P_{spin}/P_{orb}$ respectively for all hard X-ray detected mCVs used in this work. In red, we show all mCVs seen by IBIS, in blue, BAT-detected mCVs and in black, we show the only IP used in this work observed by HXD, AE Aqr. In order to obtain hardness ratios for BAT and HXD mCVs, we have reproduced their bremsstrahlung spectra (power law for AE Aqr) using the temperatures (or photon index) provided by \cite{brunsch} and \cite{terada}, respectively. We then extracted the hardness ratio from the spectra taking into account errors\footnote{We note that normalisation constants are not required when inferring hardness ratios from single model spectra} (symmetric for all) by also reproducing the hottest and coldest spectra using the published errors for each source and computing the hardness. Before being able to add the BAT points to Fig. \ref{fig:Pspin}, \ref{fig:Porb} and \ref{fig:Psync} it is necessary to remove systematic differences between the IBIS instrumental hardness and the BAT flux hardness. Fig. \ref{fig:crosscalib} shows the BAT hardness vs. the IBIS hardness for a sample of 13 IPs in common to both (we decided to exclude IGR J00234$-$6141 due to the extremely large errors in \cite{brunsch}). In red we display the one-to-one line where most of the data should sit in the absence of systematic differences between the IBIS and BAT calibrations. However, it is easily noticeable that the BAT extrapolated hardness's are systematically harder than the IBIS ones. We compensate approximately for this by fitting a straight line through the datapoints and the origin (black line in Fig. \ref{fig:crosscalib}). We then apply a correction to all the BAT points before plotting them in Fig. \ref{fig:Pspin}, \ref{fig:Porb} and \ref{fig:Psync}. We note the large scatter about the fit, which might be caused by intrinsic variability of the sources not being observed at the same time. This is also taken into account by folding the fit error into the BAT datapoints as well.

\begin{figure*}
\centering
\includegraphics[width=\textwidth, height=7cm]{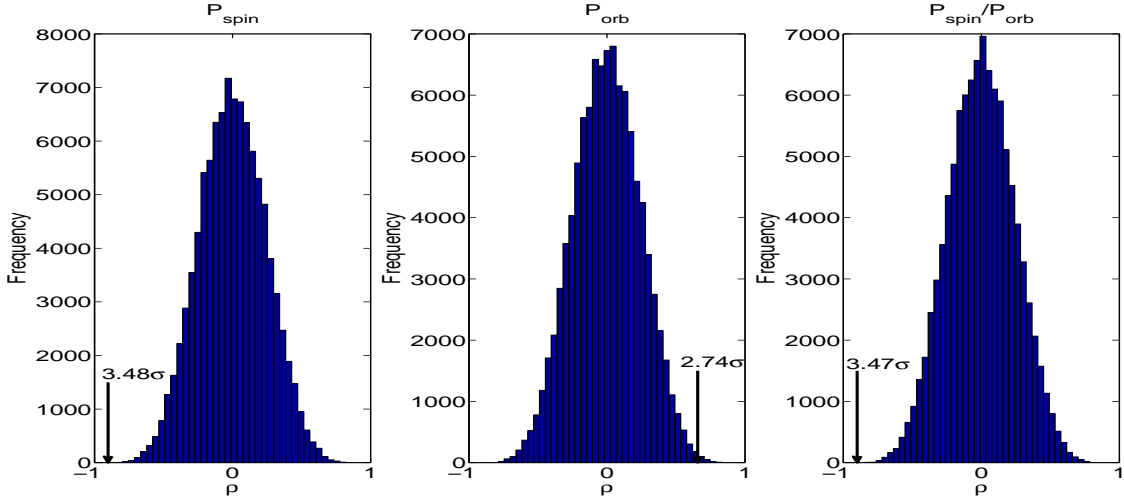}	
\caption{Results from the Monte-Carlo simulation for estimating the correlation significances of the IBIS IPs only. The distributions display the calculated $\rho$ coefficients for our mock datasets. We display with an arrow the calculated coefficient for the real set.}
\label{fig:MC2}
\end{figure*}

All three plots show evident signs of correlations with hardness ratio when considering IPs alone. In particular, when considering IPs with $P_{spin}/P_{orb}<0.1$ (systems which are on their way to equilibrium at $P_{spin}/P_{orb}=0.1$ and are therefore exclusively in the propeller stage), then the correlations become even more evident. In order to obtain a significance for these correlations observed for the IPs thought to be in the propellor stage, we performed a Spearman rank test using a Monte-Carlo scheme. At first we decided to test the IBIS observations only, as we believe these are the measurements with lowest systematic errors. For example, in order to test if the correlation between hardness and $P_{spin}$ is significant, we created 100,000 mock data sets containing the same number of points but shuffling the $P_{spin}$ values randomly each time. Moreover, in order to take the hardness uncertainties into account, we replaced each hardness value with a random variable drawn from a normal distribution whose mean is equal to the observed hardness and whose standard deviation is equal to the error on the observation. We then calculated the Spearman rank coefficients, $\rho$. The distributions of the coefficients for all $P_{spin}$, $P_{orb}$ and $P_{spin}/P_{orb}$ are shown in Fig. \ref{fig:MC2}. Also displayed in each panel is the significance for the calculated Spearman rank value of the real data. These are $3.48\sigma$, $2.74\sigma$ and $3.47\sigma$ for $P_{spin}$, $P_{orb}$ and $P_{spin}/P_{orb}$ respectively.

\begin{figure*}
\centering
\includegraphics[width=\textwidth, height=7cm]{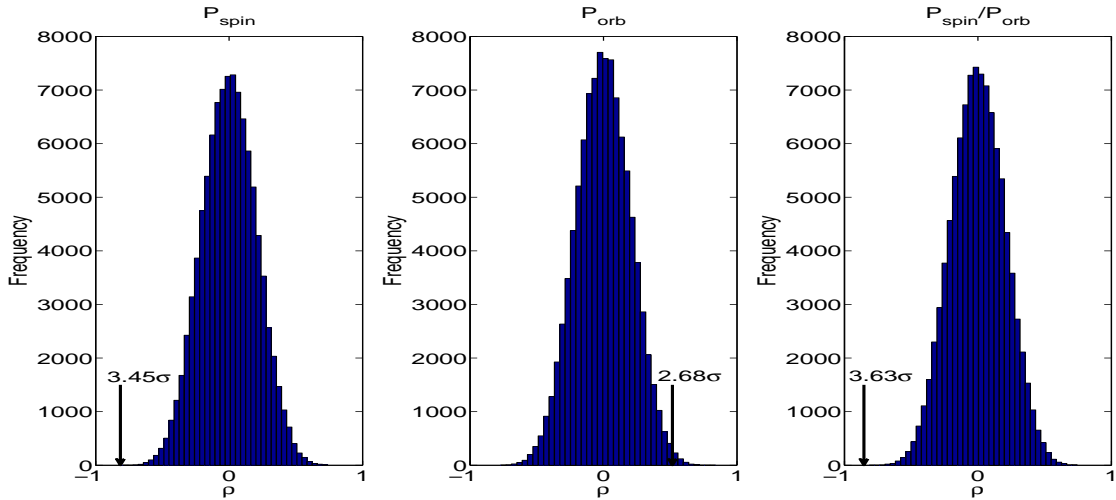}	
\caption{Results from the Monte-Carlo simulation for estimating the correlation significances of the IBIS, BAT and HXD IPs. The distributions display the calculated $\rho$ coefficients for our mock datasets. We display with an arrow the calculated coefficient for the real set.}
\label{fig:MC3}
\end{figure*}

The fact that the correlation is apparent in all three plots is somewhat expected, since $P_{spin}$ and $P_{orb}$ are not independent, but expected to evolve together (\citealt{norton_theo1,norton_theo2}). At this stage we perform the same exercise as for Fig. \ref{fig:MC2}, but this time including the additional IPs above the period gap observed by BAT only, and AE Aqr as observed by HXD. The results for this simulation are presented in Fig.\ref{fig:MC3}. 

\begin{figure*}
\centering
\includegraphics[width=\textwidth, height=12cm]{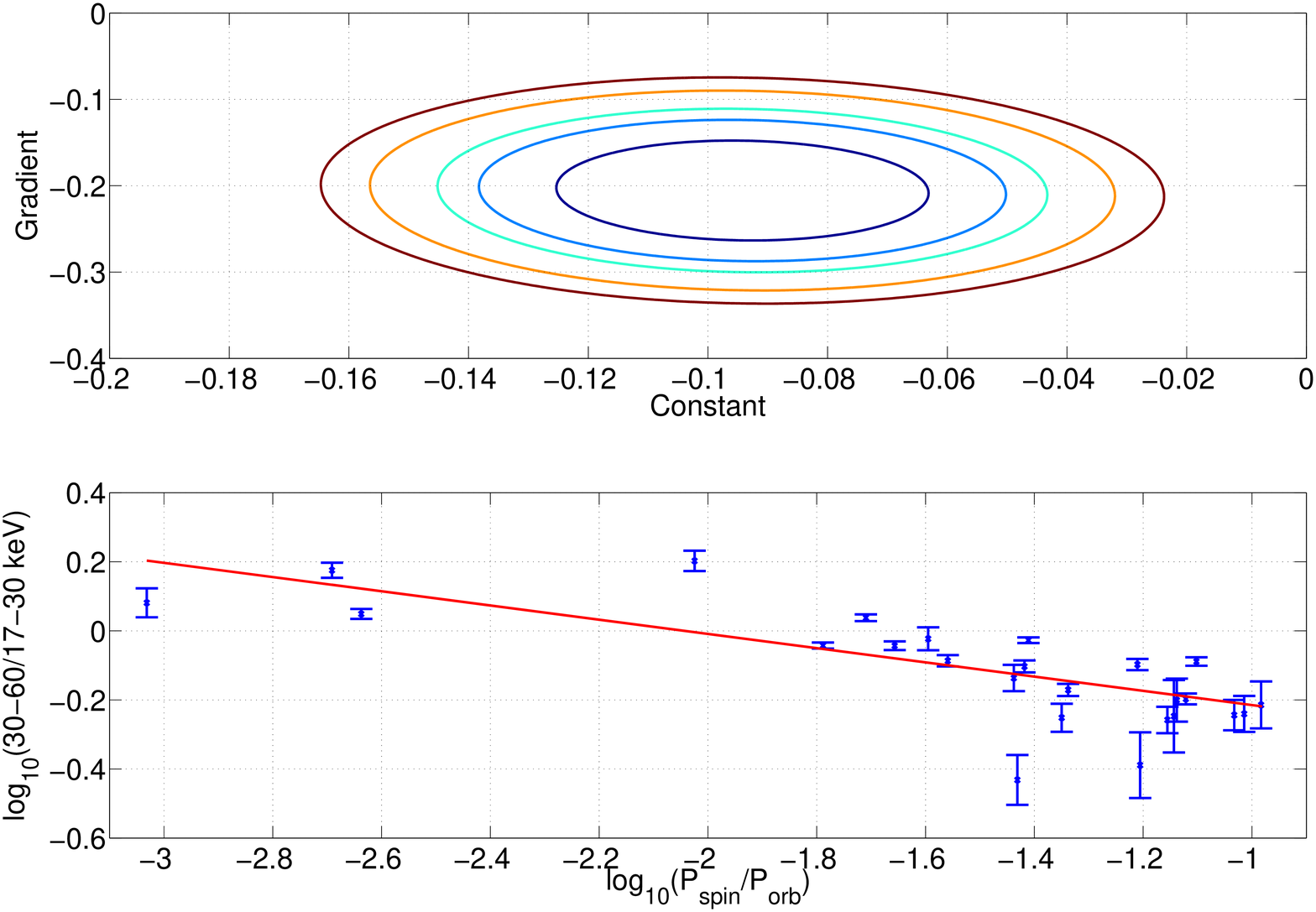}	
\caption{Top panel: Contours for a linear fit to the datapoints in the bottom panel. 
Lines display contours for $\sigma=$ 1, 2, 3, 4, 5. Bottom panel: IBIS IP hardness as a function of synchronicity. The equation resulting from the fit is 30-60/17-30 keV = $(({P_{spin}/P_{orb}})-0.0259)^{-0.21\pm0.05} - 10^{0.09\pm0.03}$ }
\label{fig:cont}
\end{figure*}

As a final step we decided to extend our analysis further for the observed correlation in $P_{spin}/P_{orb}$ vs. hardness, given that, from an evolutionary perspective, it is expected to be the most relevant parameter (\citealt{norton_theo1,norton_theo2}). We produce a linear fit in log-log space to the $P_{spin}/P_{orb}$ vs. hardness plot in Fig. \ref{fig:Psync} for the hard X-ray selected mCVs used in this work\footnote{We have tried various polynomials but these all worsened the fit}. When this is done, a non acceptable value of 4.5 is obtained for a reduced $\chi^2$. However, following a similar procedure to \cite{tremaine} and \cite{mchardy}, we introduce a small intrinsic dispersion of 0.009 to all our datapoints in linear space. This corresponds to $\approx 1\%$ for the soft IPs in Fig. \ref{fig:Psync} and $\approx 0.5\%$ to the hardest IPs. There may be many reasons why such a small error might be introduced, ranging from a $\approx 1\%$ error in the IBIS response, to any small spectral variability intrinsic to the observed systems. The addition of this intrinsic dispersion lowers the reduced chi-squared to unity and results in more conservative errors on the fit parameters. In Fig.\ref{fig:cont} we display the contour plots for our linear fit in log-log space in the top panel, and the fit itself on the bottom. Note that the contours represent lines of $\sigma=$1, 2, 3, 4, 5. Again one can see that a simple constant value straight line fit is not consistent with the data. The resulting equation to the fit can be expressed as 30-60/17-30 keV = $(({P_{spin}/P_{orb}})-0.0259)^{-0.21\pm0.05} - 10^{0.09\pm0.03}$, and may prove useful for modeling these systems and observations in the future. However at this stage, this is a purely empirical model.

\section{Discussion}
IBIS has so far detected 32 CVs (23 spatially correlated with known CVs + 9 new, optically confirmed discoveries). The majority are intermediate polars, but IBIS has also detected the bright dwarf nova SS Cyg and a few polars. This sample is an extension of the previously presented sample of IBIS CVs by \cite{barlow}, which also showed that the spectral characteristics of these objects in the 20-100 keV range are actually quite similar and compare well with previous high-energy spectral fits. Moreover BAT has observed a similar number of objects (with many in common), which shows that modern hard X-ray surveys are able to consistently observe mCVs at higher energies than before. 
 
The high incidence of IPs in our sample is not unexpected. \cite{kuijpers} and many others, including \cite{warner}, have suggested that the lower levels of hard X-rays/soft gamma-rays emission from polars may well be related to the low accretion rate and stronger magnetic fields compared to IPs. It has also been suggested that the strong magnetic fields in polars are able to produce a more ``blobby'' flow (\citealt{frank}). These high density ``blobs'' are then able to penetrate deeper within the post-shock region and contribute more to the blackbody component of the broadband X-ray spectrum, and less to the bremsstrahlung component, making the broadband X-ray spectrum of IPs harder, and hence more luminous in the hard X-rays.

Two out of four known APs are observed in our sample. The remaining two happen to be in a region of low IBIS exposure and are likely more distant. \cite{schwarz} have already shown that BY Cam, one of the observed APs, has different properties from those of normal polars, in particular having a much higher accretion rate. We therefore tentatively conclude that IBIS has not yet seen the two missing APs due to their distance/exposure. Our sample only includes 2 definite synchronous polars, and we do not expect many of these systems to be observed in the future with higher sensitivities above $17$ keV.        

Of the many observational characteristics presented here, one feature in the $P_{spin}-P_{orb}$ plane has stood out since the first study by \cite{barlow}: a very low number of IPs below the period gap are detected with hard X-ray telescopes. The only exception in our study is the very nearby IP EX Hya. We can compare this result to the theoretical models of \cite{norton_theo1, norton_theo2}. It is believed that the IPs below the period gap have accretion flows which are very different to the IPs above the period gap. In particular systems below the period gap with high $P_{spin}/P_{orb}$ such as EX Hya display ring-like accretion. We would therefore not necessarily expect these systems to behave in the same way as the systems above the period gap, and, in particular, we would not necessarily expect them to emit such high energy photons. This is still an open question, and, as mentioned by \cite{norton_conf}, only deeper hard X-ray exposures will reveal if this subclass of IPs displaying ring-like accretion is able to produce similar amounts of hard X-rays as those observed in disk-fed systems at longer orbital periods. It has already been mentioned (\citealt{norton_theo1}) that as mCVs evolve through the period gap the magnetic field of the WD may be able to resurface when accretion stops. This can allow the system to synchronise, and we would then expect a system jumping from $P_{spin} \approx 0.1P_{orb}$ above the period gap to $P_{spin} \approx P_{orb}$ below the period gap. Moreover we add to this that any system with $P_{spin}/P_{orb} \geq 0.6$ will never reach equilibrium until it reaches total synchronisation (\citealt{norton_theo1,norton_theo2}). 

All of the hard X-ray detected IP systems display $P_{spin}<0.1P_{orb}$. This may be more observational evidence for different kinds of accretion flows within the IP class, supporting the models of \cite{norton_theo1,norton_theo2}. More evidence for these models comes from the non-detection of objects in any wavelength range within the synchronicity gap (a region above the period gap within $P_{spin}/P_{orb}>0.3$ and $P_{spin}/P_{orb}<1$). Such systems are predicted not to exist since IPs tend to evolve within the $P_{spin}-P_{orb}$ plane towards their equilibrium spin rate at $P_{spin}\approx0.1P_{orb}$ above the period gap or at $P_{spin} \approx 0.6P_{orb}$ below the gap. \cite{norton_theo1, norton_theo2} have predicted that low synchronisation mCVs have propeller accretion flows and are all trying to reach equilibrium moving towards $P_{spin} \approx 0.1P_{orb}$. This equilibrium arises due to the WD trying to balance angular momentum with the surrounding blobs. We believe this is the case for most hard X-ray selected IPs, as relatively few have yet been found above $P_{spin}=0.1P_{orb}$ where the accretion flow is thought to take the form of a stream. 

Perhaps the most interesting result of this study is the discovery of a correlation between 30-60/17-30 keV hardness and spin/orbital parameters for IPs. No similar correlation has been reported before, probably because previously measured X-ray hardness ratios of IPs were generally restricted to the range of approximately $\sim 0.5 - 10$~keV. Such ratios sample the lower end of the bremsstrahlung spectrum and the upper end of the blackbody spectrum, without fully measuring either component, while X-rays below $\sim 1-2$~keV are more easily affected by photoelectric absorption caused by the intervening accretion curtains. In contrast, we note that the spectral hardness variations we have measured in our hard X-ray detected IPs span the energy range $\sim 17 - 60$~keV and are {\em only} sampling the bremsstrahlung component of the spectrum. Therefore these observations tell us nothing about the relative contributions of the bremsstrahlung component emitted by the cooling plasma below the accretion shock and the blackbody component arising from the heated WD surface. Instead, they are directly sampling the relative contributions of the multi-temperature bremsstrahlung components that arise in the plasma below the shock front (given that the plasma cools as it settles towards the WD surface).

In order to explore a possible reason for the correlation between X-ray spectral hardness and spin-to-orbital period ratio in the hard X-ray detected IPs, we begin with the uncontroversial idea that the WDs in IPs are mostly accreting close to their equilibrium spin rates (\citealt{norton_theo1}). Hence, their spin-to-orbital period ratios are an indication of their magnetic field strength (see Fig. 6 of \citealt{norton_theo1}). Broadly speaking, for smaller $P_{spin}/P_{orb}$, the surface magnetic field strength is smaller. So these systems will have smaller magnetospheric radii, and the material will attach onto field lines closer to the WD. This will give rise to a larger footprint area in those systems with smaller values of $P_{spin}/P_{orb}$. That is to say, faster spinning WDs will have larger accretion footprints beneath a wide accretion curtain. Evidence for this also comes from the observed double-peaked pulse profiles observed in fast spinning WD (hence having a small magnetic field) as described in \cite{norton99}, resulting from the optical depths across and along the accretion curtains as the WD rotates. Although the conventional theory of accretion shocks suggests that wider shocks will be taller, we here investigate the consequences if the opposite is true.

By spreading the material over a larger area beneath a shock of relatively low height, we suggest that the resulting bremsstrahlung X-ray emission may have a harder spectrum, possibly because the accretion is closer to the WD surface and there is less distance for the plasma to travel as it cools within the post-shock region towards the WD surface and so there is less contribution from cooler bremsstrahlung components. In contrast, the systems with a relatively slowly spinning WD have a larger  $P_{spin}/P_{orb}$ value, so their magnetic field strength is larger, their magnetospheric radius is larger, and their accretion footprint is smaller. We suggest that in such cases, the X-ray emitting region may sit beneath a tall but narrow accretion curtain, and then this geometry gives rise to a softer bremsstrahlung spectrum, possibly because the accretion shock is further from the WD surface and so there is a greater distance for the plasma to travel within the post-shock region and cool as it falls towards the WD surface. This interpretation also helps explain the low detection number in the hard X-ray domain of EX Hya-like systems below the period gap with high $P_{spin}/P_{orb}$ which are thought to display ring-like accretion. In these cases the magnetospheric radius extends to a very large distance from the WD implying a very small footprint area. If this then means a very tall shock height (in line with the interpretation above), then the plasma will have a long distance over which to cool as it travels towards the surface, and the spectrum will be dominated by softer photons.

Contrary to this interpretation are the ideas presented by \cite{suleimanov} and \cite{brunsch}, where the spectral hardness in the hard X-ray range could be used in order to obtain the WD mass. However it is not yet clear if this idea could explain the hardness correlations presented here. In particular if the hard X-ray spectral hardness could be used to infer the mass of the WD primary, then we would also exepect the WD mass to correlate with $P_{spin}$, $P_{orb}$ and $P_{spin}/P_{orb}$, and this is not the case. Moreover the best evidence supporting this idea is presented in Figure 6 from \cite{suleimanov} and Figure 4 from \cite{brunsch}, where the estimated WD masses inferred from the hard X-rays are plotted against the estimated WD masses inferred from other methods. It is clear by looking at the plot that large scatter is present and that the error bars are quite large. We thus believe that no firm conclusion on the connection between WD masses and spectral hardness can be obtained at the moment.

As far as we are aware, some detailed modelling of accretion shocks has been performed by \cite{canalle} and \cite{saxton}, however their model X-ray spectra obtained by the simulations only extends to 10 keV, which is not high enough for the data considered here. We point out that no one has yet modelled whether the multi-temperature bremsstrahlung spectrum is different for a wide, low accretion curtain compared with a tall, narrow accretion curtain, but we suggest this would be a worthwhile test to carry out. 

We recognise that our suggestion of wide/low accretion regions in the case of rapidly spinning white dwarfs, and narrow/tall accretion regions in the case of slowly spinning white dwarfs is contrary to the prediction of standard shock theory that wider accretion regions are taller. However, there is much in the detailed accretion flows of magnetic CVs that remains poorly understood and we put our suggestion forward in the spirit of advancing understanding of the subject. It then provides a possible way of understanding the hardness - period correlation observed here which would otherwise be unexplained.

\section{Conclusions} 

This paper has presented a catalogue and analysis of a sample of CVs detected in the hard X-ray range ($>$ 17 keV) with IBIS, BAT and HXD. As with previously compiled high-energy samples of CVs, it is shown that most systems are magnetic. Moreover, some of the detected systems are very rare types of objects (2 APs). The sample is dominated by intermediate polars, with only 2 synchronous polars. This suggests the broadband X-ray/gamma-ray spectrum of IPs is harder and more luminous than that of polars. This could be the effect of accretion rate and magnetic field strength, where IPs have higher accretion and weaker magnetic fields relative to polars, depositing a somewhat smoother accretion stream onto the poles. By contrast to the polars, the accretion flow in IPs may therefore not bury itself deep within the post-shock region, producing a harder broadband X-ray/gamma-ray spectrum.

We have shown that only IPs with $P_{spin}/P_{orb}<0.1$ are consistently found by hard X-ray surveys. Moreover, we have examined the observational properties of mCVs in the $P_{orb} - P_{spin}$ plane. We find that the observations are consistent with the theoretical models of \cite{norton_theo1,norton_theo2} for mCV evolution, where IPs tend to cluster at about $P_{spin}/P_{orb} \approx 0.1$, and none have yet been observed in the hard X-ray regime above $P_{spin}/P_{orb} \approx 0.1$. Also observed and predicted is the observation that a very low number of IPs are found in any wavelength range within the synchronicity gap: a region between $P_{spin}/P_{orb} \approx 0.3$ and $P_{spin}/P_{orb}=1$.

The paper has also presented the first observed correlations between 30-60 keV/17-30 keV hardness and $P_{spin}$, $P_{orb}$ and $P_{spin}/P_{orb}$. The correlations have been statistically tested using Monte Carlo simulations. 

In an attempt to explain our result we have suggested that hard X-ray selected IPs are spinning towards their equilibrium, so that their spin period is an indicator of magnetic field strength. This in turn will give rise to a short, but wide, post-shock region for fast spinning WDs (and therefore possessing a relatively weak magnetic field) making their hard X-ray spectra harder. In contrast slowly spinning WDs will have a tall but narrow post-shock region (possessing a relatively high magnetic field), yielding a cooler bremsstrahlung component in the hard X-rays. 

All of the observations presented in this paper are consistent with mCV evolution models. It is very likely that hard X-ray missions will continue to increase this sample of mCVs, and it is also expected that more unidentified hard X-ray sources will be identified as IPs with more optical follow-ups. In particular, more observations will allow us to establish if any IPs are detected below the period gap and if any IPs will ever get detected in the IBIS energy range above $P_{spin}/P_{orb} \approx 0.1$ in order to establish if the Norton accretion models are a plausible explanation to the observed systems. This also implies that hard X-ray selected samples could have their own biases, however more analysis will have consequences on evolution studies of these exotic magnetic systems.

\section*{Acknowledgements}
We acknowledge the following funding by STFC grant PPA/S/S/2006/04459 and PP/C000714/1 and ASI-INAF grant I008/07/0. 
We are also grateful to Margaretha Pretorius at SAAO for providing spin periods for some IPs before publishing and Prof. Brian Warner for useful discussions. Moreover we are very thankful to the anonymous referee for useful and constructive feedback. This research has made use of NASA’s Astrophysics Data System Bibliographic Services, of the SIMBAD database, operated at CDS, Strasbourg, France, as well as of the NASA/IPAC Extragalactic Database (NED), which is operated by the Jet Propulsion Laboratory, California Institute of Technology, under contract with the National Aeronautics and Space Administration. 

\bibliographystyle{mn2e}
\bibliography{ibis_cvs}

\label{lastpage}

\end{document}